# Interfacial Thermal Conductance Spectrum in Nonequilibrium Molecular Dynamics Simulations Considering Anharmonicity, Non-homogeneity and Quantum Effects

Yixin Xu[1], Lina Yang[2*] and Yanguang Zhou[1*]

[1]*Department of Mechanical and Aerospace Engineering, The Hong Kong University of Science and Technology, Clear Water Bay, Kowloon, Hong Kong SAR*

[2]*School of Aerospace Engineering, Beijing Institute of Technology, Beijing, 100081, China*

## Abstract

Interfacial thermal transport is critical for many thermal-related applications such as heat dissipation in electronics. While the total interfacial thermal conductance (ITC) can be easily measured or calculated, the ITC spectral mapping has been investigated only recently and is not fully understood. By combining nonequilibrium molecular dynamics simulations and atomistic Green's function method, we systematically investigate the ITC spectrum across an ideal interface, i.e., the argon-heavy argon interface. Our results show that the ITC spectrum increases gradually with temperature as more phonons and anharmonic scattering channels are activated, e.g., the vibrations with frequencies larger than 1 THz can contribute 5% (15%) to the total ITC at 2 K (40 K) through anharmonic phonon scatterings channels. We further find that the ITC spectrum from the left interfacial Hamiltonian is quite different from that of the right interfacial Hamiltonian, which stems from the asymmetry of anharmonic phonon scatterings caused by the dissimilar vibrational property of the two interfacial contacts. While all the phonons are involved in the

---

Author to whom all correspondence should be addressed. Email: maeygzhou@ust.hk (Y. Zhou), yangln@bit.edu.cn (L. Yang)

anharmonic scatterings for the heavy argon interfacial Hamiltonian, these phonons involved in the anharmonic phonon scatterings from the argon interfacial Hamiltonians are mainly these vibrations with frequency smaller than 1 THz (i.e., the cut-off frequency of heavy argon). Finally, we find the quantum effect is important for the ITC spectrum at low temperatures, e.g., below 30 K in our systems. Our results here systematically investigate the influence of anharmonicity, non-homogeneity, and quantum effects on the ITC spectrum, which is critical for designing and optimizing the interfaces with better performance.

# I. INTRODUCTION

Interfacial thermal transport is critical for heat dissipation in electronics and the energy conversion in thermoelectrics. Quantifying the interfacial thermal conductance (ITC) spectrum at interfaces is quite important for investigating and manipulating interfacial thermal transport. The ITC spectrum has been widely investigated, mainly through empirical models [1,2] and atomistic Green's function (AGF) method [3–5], and permutations thereof that only consider the elastic scatterings after Kapitza firstly defined the interfacial heat transfer between two different materials in 1941 [6]. Only recently, the anharmonic phonon scatterings are introduced into the analysis of the ITC spectrum. For instance, Guo *et al.* [7,8], and Dai and Tian [9] have included the three-phonon scatterings in their AGF calculated ITC spectrum. Kimmo *et al.* [10,11], Zhou *et al.* [12–14], and Xu *et al.* [15] include the anharmonic phonon scatterings in their ITC spectrums in the frame of non-equilibrium molecular dynamics (NEMD). In the equilibrium molecular dynamics (EMD) framework, Gordiz and Henry [16–18] calculate the ITC spectrum in which all anharmonic phonon scatterings are inherently included. Although tremendous progress in quantifying the ITC spectrum at interfaces has been achieved, a systematic investigation of the influence of anharmonicity, non-homogeneity, and quantum effects at the interface on the ITC spectrum is still lacking.

In this paper, we calculate the ITC spectrum across an ideal interface, i.e., Ar-heavy Ar interfaces, by applying the interatomic forces in the spectral heat current decomposition in the frame of NEMD [10,12]. The anharmonicity of the ITC spectrum is then investigated by comparing it with the harmonic ITC spectrum calculated using AGF or second-order force constant based spectral heat current method at extremely low temperature. Our results show that the ITC spectrum increases gradually with temperature as more phonons and anharmonic scattering channels are activated, and the vibrations with frequencies larger than 1 THz can contribute 15% to the total ITC at 40 K through anharmonic scatterings channels. We further find that the spectral ITC from the interfacial Hamiltonians of the argon side is quite different from that of the heavy argon side. While all the phonons from the heavy argon interfacial Hamiltonians are found to be involved in the ITC, these phonons with frequency larger than 1 THz from the argon Hamiltonians contribute a little, e.g., 5% (4% after quantum correction) at 30 K, to the total thermal conductance. The asymmetry interfacial phonon energy of two contacting interfacial sides resulting from the anharmonic effects may be used to manipulate the heat flow across the interface [19,20], which may benefit the design of thermal logic devices.

The paper is organized as follows. In Sec. II, we introduce the theory to calculate the ITC spectrum for the two contact parts considering elastic phonon scatterings [Eq. (22)] using second-order force constants, and all the high-order phonon scatterings [Eq. (15)] using fully expressed interatomic forces. We also briefly introduce the AGF, which will be used to validate our calculated harmonic ITC spectrum. The computational details, including model construction, atomic potentials, and the test of the size effects, are presented in Sec. IIC. In Sec. IIIA and IIIB, we investigate how the anharmonicity and non-homogeneity at the interface will affect the ITC spectrum. How the anharmonicity will affect the non-homogeneity in the ITC spectrum is then discussed in Sec. IIIC. Before closing, we discuss the quantum effect caused by the Boltzmann distribution in molecular dynamics on the ITC in Sec. IIIC. We present the conclusions in In Sec. IV.

**II. THEORY**

## A. Spectral interfacial thermal conductance

The interfacial heat current is defined as the net energy exchange rate from the left side to the right side across the target interface between two materials. If we take the positive sign for the total interfacial heat flow, the heat current from the left to the right side $Q_{Left \to Right}$ can be expressed as [17,21]

$$Q_{Left \to Right} = \frac{\partial H_{Right}}{\partial t} - \frac{\partial H_{Left}}{\partial t} \qquad (1)$$

where $H_{Left}$ and $H_{Right}$ are the Hamiltonians of the left and right interfacial region, respectively. In our calculations, the left (right) interfacial region is defined as one unit cell distance from the interface. The Hamiltonian can be written as

$$H = \sum_i^N (\frac{1}{2} m_i \vec{v}_i^{\,2} + \Phi_i) \qquad (2)$$

in which $\Phi_i(\vec{r}_1, \vec{r}_2, \ldots \vec{r}_n)$ is the potential energy of atom $i$, $\vec{r}_i$ is the atomic position and the potential energy can be written in the form of

$$\Phi_i = \frac{1}{2} \sum_{j \neq i}^N U_{ij}(\vec{r}_{ij}) \qquad (3)$$

where $\vec{r}_{ij} \equiv \vec{r}_j - \vec{r}_i$ and $U_{ij}$ are the distance and potential between atom $i$ and atom $j$, respectively. Therefore, the energy exchange rate in Eq. (1) can be rewritten as

$$\frac{\partial H}{\partial t} = \sum_i^N (m_i \vec{v}_i \frac{d\vec{v}_i}{dt} + \frac{\partial \Phi_i}{\partial \vec{r}_i} \frac{\partial \vec{r}_i}{\partial t}) \qquad (4)$$

In a solid material, it is known that the convection heat current, i.e., the first term in Eq. (4), caused by the atomic diffusion can be ignored. Then, the interfacial net energy exchange rate, i.e., the interfacial heat current, contributed from the left interfacial Hamiltonian has the formula of

$$\frac{\partial H_{Left}}{\partial t} = \sum_{i \in Left}^{N_{Left}} \left( \sum_{i' \in Left}^{N_{Left}} \frac{\partial \Phi_i}{\partial \vec{r}_{i'}} \frac{\partial \vec{r}_{i'}}{\partial t} + \sum_{j \in Right}^{N_{Right}} \frac{\partial \Phi_i}{\partial \vec{r}_j} \frac{\partial \vec{r}_j}{\partial t} \right) \qquad (5)$$

and the heat current contribution from the right interfacial Hamiltonian is

$$\frac{\partial H_{Right}}{\partial t} = \sum_{j \in Right}^{N_{Right}} \left( \sum_{i \in Left}^{N_{Left}} \frac{\partial \Phi_j}{\partial \vec{r}_i} \frac{\partial \vec{r}_i}{\partial t} + \sum_{j' \in Right}^{N_{Right}} \frac{\partial \Phi_j}{\partial \vec{r}_{j'}} \frac{\partial \vec{r}_{j'}}{\partial t} \right) \qquad (6)$$

in which $\partial \vec{r}_i / \partial t$ is the velocity of atom $i$, $N_{Right}$ and $N_{Left}$ are the number of atoms of the right side and the left side, respectively. The partial differential function of atomic potential $\Phi_i(\vec{r}_1, \vec{r}_2, \ldots \vec{r}_n)$ equals to

$$\frac{\partial \Phi_i}{\partial \vec{r}_{i'}} = \frac{1}{2} \frac{\partial U_{ii'}}{\partial \vec{r}_{i'}} = \frac{1}{2} \vec{F}_{ii'} \qquad (7)$$

$$\frac{\partial \Phi_i}{\partial \vec{r}_j} = \frac{1}{2} \frac{\partial U_{ij}}{\partial \vec{r}_j} = \frac{1}{2} \vec{F}_{ij} \qquad (8)$$

where $\vec{F}_{ii'}$ refers to the force exerted on atom $i$ by atom $i'$ from the same interfacial side, $\vec{F}_{ij}$ is the atomic force across the interface. In our calculations, the atomic forces are calculated and updated based on the time-dependent atomic configurations. Therefore, the interfacial heat current from the left interfacial side to the right interfacial side can be written as

$$Q_{Left \to Right} = \frac{1}{2} \underbrace{\left[ \sum_{i \in Left}^{N_{Left}} \sum_{j \in Right}^{N_{Right}} \langle \vec{F}_{ji} \vec{v}_i \rangle + \sum_{j < j' \in Right}^{N_{Right}} \langle \vec{F}_{jj'} (\vec{v}_{j'} - \vec{v}_j) \rangle \right]}_{Right-H} - \frac{1}{2} \underbrace{\left[ \sum_{i \in Left}^{N_{Left}} \sum_{j \in Right}^{N_{Right}} \langle \vec{F}_{ij} \vec{v}_j \rangle + \sum_{i < i' \in Left}^{N_{Left}} \langle \vec{F}_{ii'} (\vec{v}_{i'} - \vec{v}_i) \rangle \right]}_{Left-H} \qquad (9)$$

where the $\langle\ \rangle$ denotes the time average in MD simulations, and the Left-H (Right-H) refers to the interfacial heat current contributed by the interfacial left (right) Hamiltonian. The second term of the left or right Hamiltonian resulting heat current in Eq. (9) describes the energy exchange rate on the same sides and can be ignored because of the ergodicity of atomic velocities terms. The interfacial heat current is then simplified into

$$Q_{Left \to Right} = -\frac{1}{2}\sum_{i \in Left}^{N_{Left}} \sum_{j \in Right}^{N_{Right}} [\underbrace{\langle \vec{F}_{ij}\vec{v}_j \rangle}_{Left\text{-}H} + \underbrace{\langle \vec{F}_{ij}\vec{v}_i \rangle}_{Right-H}] \qquad (10)$$

At the same time, we can define the heat current auxiliary correlation function $C_{ij}(\tau)$ from atom $i$ to $j$ [10,12]

$$C_{ij}(\tau) = -\frac{1}{2}\langle \vec{F}_{ij}(\tau)\vec{v}_i(0) + \vec{F}_{ij}(\tau)\vec{v}_j(0) \rangle \qquad (11)$$

in which, $\tau$ is the correlation time between the interatomic forces and atomic velocity. The Fourier's transform of $C_{ij}(\tau)$ is then given by

$$\tilde{C}_{ij}(\omega) = \int_{-\infty}^{+\infty} C_{ij}(\tau)e^{-i\omega\tau}d\tau \qquad (12)$$

The real and imaginary parts of $\tilde{C}_{ij}(\omega)$ in Eq. (12) are even and odd functions since the $C_{ij}(\tau)$ is real. Correspondingly, $C_{ij}(\tau)$ can be then rewritten as

$$C_{ij}(\tau) = \int_{-\infty}^{+\infty} \tilde{C}_{ij}(\omega)e^{i\omega\tau}\frac{d\omega}{2\pi} = 2\operatorname{Re}\left[\int_{0}^{+\infty} \tilde{C}_{ij}(\omega)e^{i\omega\tau}\frac{d\omega}{2\pi}\right] \qquad (13)$$

Therefore, we can also rewrite the heat current from atom $i$ to atom $j$ in the form of

$$Q_{ij} = \lim_{\tau \to 0} C_{ij}(\tau) = 2\operatorname{Re}\left[\int_{0}^{+\infty} \tilde{C}_{ij}(\omega)e^{i\omega\tau}\frac{d\omega}{2\pi}\right] = 2\operatorname{Re}\left[\int_{0}^{+\infty}\int_{-\infty}^{+\infty} C_{ij}(\tau)e^{-i\omega\tau}d\tau \frac{d\omega}{2\pi}\right] \qquad (14)$$

Combined with Eq. (10), the ITC spectrum can be given by

$$G_{Left \to Right}(\omega) = \frac{Q_{Left \to Right}(\omega)}{A\Delta T} = \frac{2}{A\Delta T}\sum_{i \in Left}^{N_{Left}}\sum_{j \in Right}^{N_{Right}} \operatorname{Re}\left[\int_{-\infty}^{+\infty} C_{ij}(\tau)e^{-i\omega\tau}d\tau\right] \qquad (15)$$

where $A$ is the area of the interface and $\Delta T$ is the temperature drop at the interface.

On the other side, the total interfacial heat current is thought to be contributed from both the left and the right interfacial Hamiltonians, i.e., $\partial H_{Left}/\partial t$ and $\partial H_{Right}/\partial t$ as shown in Eq. (5) and Eq. (6), respectively. The first term in Eq. (5) denotes the energy exchange within the left side and can be neglected. Combined with Eq. (8), Eq. (5) can be expressed as

$$Q_{Left-H} = -\frac{1}{2} \sum_{i \in Left}^{N_{Left}} \sum_{j \in Right}^{N_{Right}} \left\langle \vec{F}_{ij} \vec{v}_j \right\rangle \tag{16}$$

Similarly, the heat current from the right interfacial Hamiltonian can be defined as

$$Q_{Right-H} = -\frac{1}{2} \sum_{i \in Left}^{N_{Left}} \sum_{j \in Right}^{N_{Right}} \left\langle \vec{F}_{ij} \vec{v}_i \right\rangle \tag{17}$$

It is worth noting that Eq. (16) should be identical to Eq. (17) based on the statistical mechanics [22]. Together with Eq. (14), the ITC spectrum resulting from the left or the right interfacial Hamiltonian is

$$G_{Left-H}(\omega) = \frac{Q_{Left-H}(\omega)}{A\Delta T} \sum_{i \in Left}^{N_{Left}} \sum_{j \in Right}^{N_{Right}} \text{Re}\left[\int_{-\infty}^{+\infty} \left\langle \vec{F}_{ji}(\tau)\vec{v}_j(0) \right\rangle e^{-i\omega\tau} d\tau \right] \tag{18}$$

$$G_{Right-H}(\omega) = \frac{Q_{Right-H}(\omega)}{A\Delta T} \sum_{i \in Left}^{N_{Left}} \sum_{j \in Right}^{N_{Right}} \text{Re}\left[\int_{-\infty}^{+\infty} \left\langle \vec{F}_{ji}(\tau)\vec{v}_i(0) \right\rangle e^{-i\omega\tau} d\tau \right] \tag{19}$$

On the other hand, based on the Taylor series, the atomic potential can be written in the form of

$$U_{ij} = U_e + \frac{\partial U_e}{\partial \vec{r}_e}\vec{u}_{ij} + \frac{1}{2}\frac{\partial^2 U_e}{\partial \vec{r}_e^2}\vec{u}_{ij}^{\,2} + o(\vec{u}_{ij}^{\,3}) \tag{20}$$

in which $U_e$ is the atomic potential energy at its equilibrium positions $\vec{r}_e$, and $\vec{u}_{ij}$ is the atomic relative displacement. The force between atom $i$ and atom $j$ can be then approximated by

$$\vec{F}_{ij} \approx -\frac{\partial^2 U_e}{\partial \vec{r}_e^2} \vec{u}_{ij} \tag{21}$$

Following Eq. (14), the spectral harmonic ITC [23] across the interface can be simplified to the form of

$$G_{Left \to Right}^{harmonic}(\omega) = \frac{Q_{Left \to Right}^{harmonic}(\omega)}{A \Delta T} = \sum_{i \in Left}^{N_{Left}} \sum_{j \in Right}^{N_{Right}} \text{Re}\left[ \int_{-\infty}^{+\infty} K_{ij} \left\langle \vec{u}_{ij}(\tau)\vec{v}_i(0) + \vec{u}_{ij}(\tau)\vec{v}_j(0) \right\rangle e^{-i\omega\tau} d\tau \right] \tag{22}$$

where $K_{ij} = \left.\frac{\partial^2 U_e}{\partial \vec{r}_e^2}\right|_{\vec{u}\to 0}$ is the second-order force constant for atom $i$ and $j$, which is calculated via the finite displacement method. The harmonic heat current [Eq. (22)] at 1 K, which should be low enough to ignore the anharmonic effect, is used to calculate the harmonic thermal conductance. It is worth noting that the harmonic thermal conductance calculated from MD simulations at 1 K is identical to the AGF result (see details below).

### B. Atomistic Green's function

In the framework of AGF [3,4], the system is divided into three interacting parts: the left, the right semi-infinite leads, and the target device. The motion of this system can be described by the matrix form

$$(\omega^2 \mathbf{I} - \mathbf{H})\Psi = 0 \tag{23}$$

where $\omega$ is the frequency, $\mathbf{I}$ is the identity matrix, $\Psi$ denotes the eigenvectors of the total system, and $\mathbf{H}$ is the dynamical matrix of the system. The 2nd order force constants that compose the dynamic matrix $\mathbf{H}$ is calculated by finite displacement method. The target device retard Green's function $\mathbf{G}_D^{ret}(\omega)$ is defined as

$$\mathbf{G}_D^{ret}(\omega) = (\omega^2 \mathbf{I} - \mathbf{H}_D - \Sigma_L - \Sigma_R)^{-1} \tag{24}$$

in which, $\mathbf{H}_D$ is the dynamic matrix of the device. The self-energy of the left lead $\Sigma_L = \mathbf{H}_{DL} g_L^{00} \mathbf{H}_{LD}$ and the self-energy of the right lead $\Sigma_R = \mathbf{H}_{DR} g_R^{00} \mathbf{H}_{RD}$, where $\mathbf{H}_{DL}$ and $\mathbf{H}_{DR}$ are the coupling matrixes between the device and the left and right leads, respectively. The left and the right uncoupled retarded Green's functions $g_L^{00}, g_R^{00}$ are calculated using the decimation method [24]. The spectral transmission $\Xi(\omega)$ is then calculated as

$$\Xi(\omega) = \text{Trace}[\Gamma_R(\omega) G_D^{ret}(\omega) \Gamma_L(\omega) G_D^{ret\dagger}(\omega)] \tag{25}$$

in which, $\Gamma_L = i(\Sigma_L - \Sigma_L^\dagger)$ and $\Gamma_R = i(\Sigma_R - \Sigma_R^\dagger)$.

Based on the Landauer formula [25,26], the total thermal conductance following the classical limit distribution $f_{classical\ limit} = k_b T/\hbar \omega$ which is the case in classical MD simulations can be obtained via

$$G_{AGF} = \frac{1}{2\pi A} \int_0^\infty k_b \Xi(\omega) d\omega \tag{26}$$

where $A$ is the section area and $k_b$ is the Boltzmann constant.

### C. Computational details

In this paper, an interfacial structure constructed by two dissimilar Lennard-Jones (LJ) face-cantered-cubic lattices, as shown in **Figure 1a**, is used to analyze the interfacial thermal transport process. The primary motivation for choosing the LJ argon-heavy argon interfaces as our sample systems is to make the computational cost of AGF and MD simulations acceptable, and the system size is large enough to exclude the size effects in the NEMD simulations. Meanwhile, the LJ potential also simplifies the following discussions since we do not include the three-body forces or the contributions from the optical phonons. Furthermore, the LJ potential includes strong non-linear terms and thus introduces higher-order phonon scatterings at the interfaces. Therefore, we would like to expect that our analysis here should be applicable to more complicated systems such as Si/Ge interfaces. The parameters [27] used in the LJ potential $U_{ij} = 4\varepsilon \left[ \left( \frac{\sigma}{r_{ij}} \right)^{12} - \left( \frac{\sigma}{r_{ij}} \right)^{6} \right]$ are $\varepsilon = 0.0104$ eV and $\sigma = 3.4$ Å. The cut-off distance for LJ interaction is $2.5\sigma$, and the simulation timestep is 1 fs. Periodical boundary conditions are applied to the system's lateral directions, i.e., perpendicular to the direction of heat flow **(Figure 1a)**. All simulations are firstly relaxed at NVT (constant particle number, volume and temperature) ensemble for 2 ns and then applied with heat current at NVE (constant particle number, volume and energy) ensemble for another 6 ns to obtain the steady-state (stable temperature distribution and steady heat current). Here, the heat current is generated by applying the temperature coupling to the two Lagevein thermostats, i.e., $T_{thermostats} = T \pm 0.5T$. The time-dependent atomic force and atomic velocity used in Eq. (10), Eqs. (16) and (17) are dumped every 10 steps from the last 1 ns steady state of the NEMD simulations. The control volume used to calculate interfacial atomic force and velocity is defined as one unit cell length at both sides of the interface (**Figure 1a**). All the MD simulations are performed by Large-scale Atomic Molecular Massively Parallel Simulator (LAMMPS) [28]. Once the steady heat current is formed, the temperature distribution along the direction of the heat current can be obtained (**Figure 1b**). The interfacial temperature jump $\Delta T$ is estimated by the difference between the linear fittings of the two sides' temperature profiles at interfaces.

The lattice constant $a$ in our simulations is temperature-dependent and ranges from 0.527 nm at 1 K to 0.537 nm at 40 K. It is noted that the size effect in NEMD simulations is strong when the system temperature is very low, e.g., 1 K and 2 K. To eliminate size effects in our NEMD simulations, the width and the length of our computational systems are chosen to be 10$a$ (**Figure 1c**) and 160$a$ (**Figure 1d**), respectively.

### III. RESULTS AND DISCUSSIONS

#### A. Spectral thermal conductance considering anharmonicity

In this section, we firstly compare the harmonic thermal conductance from MD [Eq. (22)] at low temperature with that calculated using AGF [Eq. (26)]. Zhou [29] has recently proved that all the vibrations are occupied in MD. Therefore, the thermal conductance of AGF should be equal to that calculated using MD when all the vibrations are assumed to be occupied in AGF [Eq. (26)], as shown in (**Figure 2a**). Our results (**Figure 2a**) clearly show that the harmonic thermal conductance spectrum calculated by the second-order force constant [Eq. (22)] at 1 K agrees quite well with that calculated using AGF [Eq. (26)], which once again validates that all the vibrations are occupied in MD. We also calculate ITC considering all high-order phonon scatterings at 1 K using Eq. (15), and find it is in accordance with the harmonic [Eq. (22)] and AGF [Eq. (26)] thermal conductance, which indicates the anharmonic phonon scatterings contribute a little to the interfacial thermal transport at 1 K. We also compare our calculated spectral thermal conductance using Eq. (15) at 10 K (**Figure 2b**) with the existing results from Ref. [10], and a general agreement between our results and the reference is found. While all the higher-order phonon scattering processes are considered in our results, only first-order anharmonic phonon scattering is included in the reference. As a result, our calculated ITC spectral at high frequency, i.e., $\omega > 1$ THz, which owes to the anharmonic phonon scattering process, is a little bit different from that of the reference.

We then turn to focus on how the temperature will affect the ITC spectrum across the argon-heavy argon interfaces. It is expected that the total ITC increases with the temperatures since more anharmonic phonon-phonon scatterings are activated. Our ITC spectrum further shows the enhancement resulting from all the vibrations in the system (**Figure 2c**). The contribution of these vibrations with frequencies larger than 1 THz, which can transport thermal energy through anharmonic phonon-phonon scatterings, is increasing(decreasing) from 5% at 2 K to 15% at 40 K (**Figure 2d**). Our previous work shows that the two- and three-phonon scattering processes account for more than 95% of the ITC when the temperature is lower than 30 K [13]. The higher-order (larger than 3) phonon scattering processes become important at higher temperatures, e.g., the higher-order process contributes ~9% to the total ITC at 40 K temperature [13]. Therefore, it is critical to include all high-order phonon scattering processes in the ITC spectrum calculations when the temperature is high, i.e., the temperature is larger than 30 K for our systems.

**B. The effect of non-homogeneity on interfacial thermal conductance spectrum**

As we discussed above, the interfacial heat current calculated using Eq. (10) can be divided into two identical parts contributed by the left and right interfacial Hamiltonians [Eqs. (16) and (17)]. Our numerical results show that the heat current of the left interfacial Hamiltonian $Q_{Left-H}(t)$ is exactly the same as that of the right interfacial Hamiltonian $Q_{Right-H}(t)$ (**Figure 3**). Meanwhile, it is known that the interfacial heat current can be calculated using both Eq. (10) and $Q_{NEMD} = \partial E_{bath}/\partial t$, in which $E_{bath}$ is externally applied energy. Our results here show the heat current computed using Eq. (10) agrees reasonably with that calculated using $Q_{NEMD} = \partial E_{bath}/\partial t$ (**Figure 3**).

We further investigate the thermal conductance spectrums contributed from the left [Eq. (18)] and right [Eq. (19)] interfacial Hamiltonians. While the ITC resulting from the left interfacial Hamiltonian is identical to that caused by the right interfacial Hamiltonian as the heat current contributed by the left interfacial Hamiltonian is same to that computed using the right interfacial Hamiltonian (**Figures 3**), the ITC spectrum resulting from the left interfacial Hamiltonian is found to be quite different from that contributed by the right interfacial Hamiltonian (**Figure 4a**). For comparison, we also calculate the harmonic thermal conductance of the argon-heavy argon interface via AGF [Eq. (26)]. These vibrations with frequencies larger than the cut-off frequency (i.e., 1THz), which should transport the thermal energy via anharmonic phonon scatterings, contribute largely(slightly) to the interfacial thermal transport (**Figure 4a**) resulting from the right(left) Hamiltonian. Meanwhile, these vibrations with frequencies smaller than 1 THz from either the left or right interfacial Hamiltonians are found to contribute differently to the total ITC when anharmonic phonon scatterings are involved, e.g., at 10 K, with 48% and 39% to total ITC, respectively (**Figure 4a**). It is worth noting that the harmonic phonon process should yield an identical ITC spectrum from the Hamiltonians of two sides because the phonons are transported elastically. However, the heat current across the interface shows strong asymmetry when anharmonic phonon scatterings are involved, which is attributed to the dissimilar vibrational property of the contact parts, e.g., the effective heat carriers in heavy argon are the vibrations with frequency lower than 1 THz whereas the argon inherently includes the high-frequency vibrational modes ($\omega >$ 1 THz).

### C. The effects of the anharmonicity on the non-homogeneity in spectral thermal conductance

We next investigate how the anharmonicity or equivalently temperature will affect the non-homogeneity in the ITC spectrum (**Figure 5**). At a low temperature of 2 K (**Figure 5a**), the ITC spectrums from left and right interfacial Hamiltonians are similar since the two-phonon scatterings dominate the interfacial heat transfer. When the temperature increases (**Figures 5b-d**), the asymmetry of the ITC spectrum contributed from the left or right interfacial Hamiltonian becomes evident. For the ITC spectrum resulting from the argon interfacial Hamiltonian, these phonons with frequencies larger than 1 THz, which cannot exist in the heavy argon, are found to contribute larger to the anharmonic interfacial heat transfer when the temperature increases. This is because more high-order, i.e., larger than and equal to three phonon-phonon scatterings at the interface can be activated with the temperature.

For the ITC spectrum resulting from the heavy argon interfacial Hamiltonian, these phonons with frequencies smaller than 1 THz are the main heat carriers, and their contribution to the total ITC is increasing with temperature. We also find these phonons resulting from the heavy argon interfacial Hamiltonian with frequencies larger than the cut-off frequency, i.e., 1 THz, contribute a little to the total ITC (**Figures 5c and d**). This may be because the nonlinearities in the interfacial potential induced by the temperature, which leads to the phonons with frequency larger than 1 THz appear in heavy argon side [13]. Consequently, the phonons with frequencies larger than 1 THz resulting from the argon (heavy argon) interfacial Hamiltonians contribute 0~5% (5%~14%) to the total thermal conductance for the temperature range considered here. Our results here prove the non-homogeneity in ITC spectrums contributed from the left and right interfacial Hamiltonians are increasing with temperature or equivalently anharmonicity, as more high-order phonon scattering processes will be activated at higher temperatures.

**D. Assessing the quantum effects in classical interfacial thermal conductance spectrum**

We emphasize that the analysis above assumes that all the vibrations are fully occupied, which is well-known in MD simulations [30]. Therefore, the thermal transport properties at low temperatures, e.g., well lower than the corresponding Debye temperature, predicted based on MD simulations is questionable [31]. To mitigate this, the quantum correction should be applied to our calculated thermal transport properties. Since all the vibrations are fully occupied in our MD calculations, the corresponding interfacial transmission can be then calculated based on Landauer theory [25]

$$\Gamma_{MD}(\omega,T) = \frac{Q_{MD}(\omega,T)}{k_b \Delta T} \tag{27}$$

in which $\Delta T$ is the temperature drop at the interface and $k_b$ refers to the Boltzmann constant. Our calculated phenomenological phonon transmission spectrum at 1 K is then compared with that of the AGF, and a small difference between them is found (**Figure 6a**). Therefore, the inaccuracy of phonon scatterings inside the near interfacial regions caused by the Boltzmann distribution of vibrations in MD simulations may have a small influence on the interfacial thermal transport, which follows the findings of Sadasivam *et al.* [32].

The results of Eq. (27) at different temperatures are also plotted in **Figure 6a**. The ITC can be then corrected quantumly in the form of

$$G_{QC\,or\,AGF} = \frac{1}{2\pi A} \int_0^\infty \hbar\omega \frac{\partial f(\omega,T)}{\partial T} [\Gamma_{MD}(\omega,T) \text{ or } \Gamma_{AGF}(\omega)] d\omega \tag{28}$$

where $A$ is the transverse area of the unit cell, $f(\omega,T)$ is the Bose-Einstein distribution, $\hbar$ is the reduced Planck's constant, and $T$ is the temperature of the system. It should be noted that $\Gamma_{AGF}(\omega)$ is temperature independent, denoting the two-phonon process transmission at all temperatures, which is almost equal to our MD phonon transmission at 1 K shown in **Figure 6a**.

At low temperatures, these high-frequency phonons, which should not be activated (**Figures 6a and 6b**), are included in MD simulations. Therefore, the corresponding total ITC calculated using MD is overestimated. For instance, the ITC at 1 K is approaching zero after applying the quantum correction to our MD results. Our corrected total thermal conductance and ITC spectrum also show the inaccuracy induced by the classical limit distribution in MD simulations rather than the Bose-Einstein distribution for phonons in the real situation can be ignored at high enough temperatures, e.g., the temperature is larger than 30 K for the argon-heavy argon systems (**Figure 6c**). Furthermore, at low temperatures, we find the high-frequency phonons ($\omega >$ 1 THz) are strongly affected by the distribution function, e.g., these phonons contribute 13% (0.5% after quantum correction) to the total ITC at 10 K in our simulations **(Figure 6c)**. Finally, We would like to emphasize that the quantum effect will also affect the high-order phonon scattering processes, e.g., the difference in the vibrational relaxation time for the quantum and classical crystalline silicon at a temperature of 200 K can be as large as 100 times [33], which may be important for interfacial thermal transport, while is challenging to be considered in MD simulations.

## IV. CONCLUSIONS

In conclusion, we systematically investigate the influence of anharmonicity, non-homogeneity and quantum effects on the ITC spectrum at the planar interfaces of LJ solids, i.e., argon-heavy argon interfaces, using the spectral heat current method in the frame of non-equilibrium molecular dynamics simulations and atomistic Green's function. The ITC is found to increase gradually with temperature as more phonons, and anharmonic scattering channels are activated, e.g., the vibrations with frequencies larger than 1 THz can contribute only 5% at 2 K while 15% at 40 K to the total ITC. Meanwhile, while the thermal conductance resulting from the left interfacial Hamiltonian is identical to that contributed by the right interfacial Hamiltonian, the corresponding ITC spectrums resulting from the left and right Hamiltonians are quite different due to the asymmetry anharmonic interfacial phonon scatterings caused by the dissimilar vibrational property of the two interfacial contacts. For the right interfacial Hamiltonian, all the phonons contribute to the anharmonic phonon scatterings. While for the left interfacial Hamiltonian, these phonons involved in the anharmonic thermal conductance are mainly these vibrations with frequency smaller than 1 THz (i.e., the cut-off frequency of heavy argon). The quantum effect is further found to be important for the ITC spectrum at low temperatures, e.g., below 30 K in our systems. Our results here provide a fundamental understanding of the ITC in molecular dynamics, and suggest a strategy to calculate the interfacial thermal conductance in all temperature ranges using molecular dynamics.

**Acknowledgments**

Y.Z. thanks the startup fund (REC20EGR14, a/c-R9246), SJTU-HKUST joint research collaboration fund (SJTU21EG09) and the Bridge Gap Fund (BGF.008.2021) from Hong Kong University of Science and Technology (HKUST).

**Data availability**

The datasets generated during and/or analyzed during this study are available from the corresponding author on reasonable request.

**Code availability**

The codes used in this study are available from the corresponding author on reasonable request

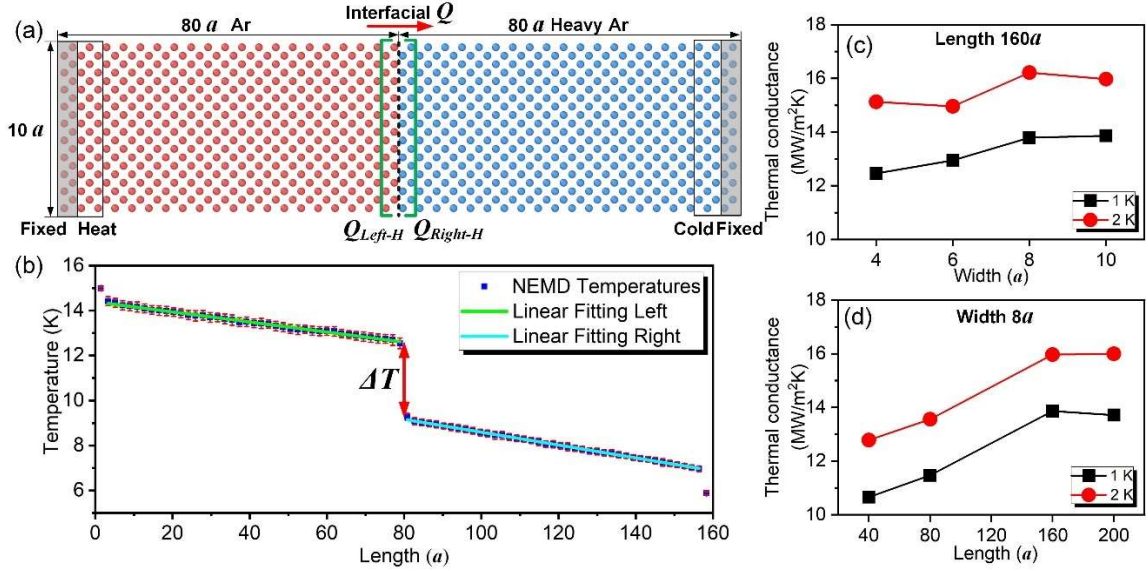

**FIG. 1** (a) The interfacial system constructed by two mass mismatched Lennard-Jones solids. The mass of the left-side argon is 39.95 amu and the mass of the right-side heavy argon is 159.8 amu. The same temperature-dependent lattice constant, i.e., $a = 0.529$ nm at 10 K and $a = 0.527$ nm at 1 K, is used in both argon and heavy argon to eliminate the lattice mismatch. Two ends of the system are fixed, and atoms belong to the heat and cold regions are coupled to Langevin thermal baths at different temperatures to generate the heat current $Q$ across the interface. We choose the atoms in the control volume region, one unit cell length at both sides of the interface illustrated by the green line, to calculate interfacial atomic force and velocity. (b) Typical temperature distribution along with the system at 10 K. The temperature jump $\Delta T$ at the interface is obtained by extrapolating the linear fittings of two sides at the contact interface. (c) The length-dependent and (d) the width-dependent ITC at various temperatures. The thermal conductance is calculated using Fourier's law, $G = Q_{NEMD}/(A\Delta T)$, where $Q_{NEMD}$ is the heat current and $A$ denotes the cross-sectional area.

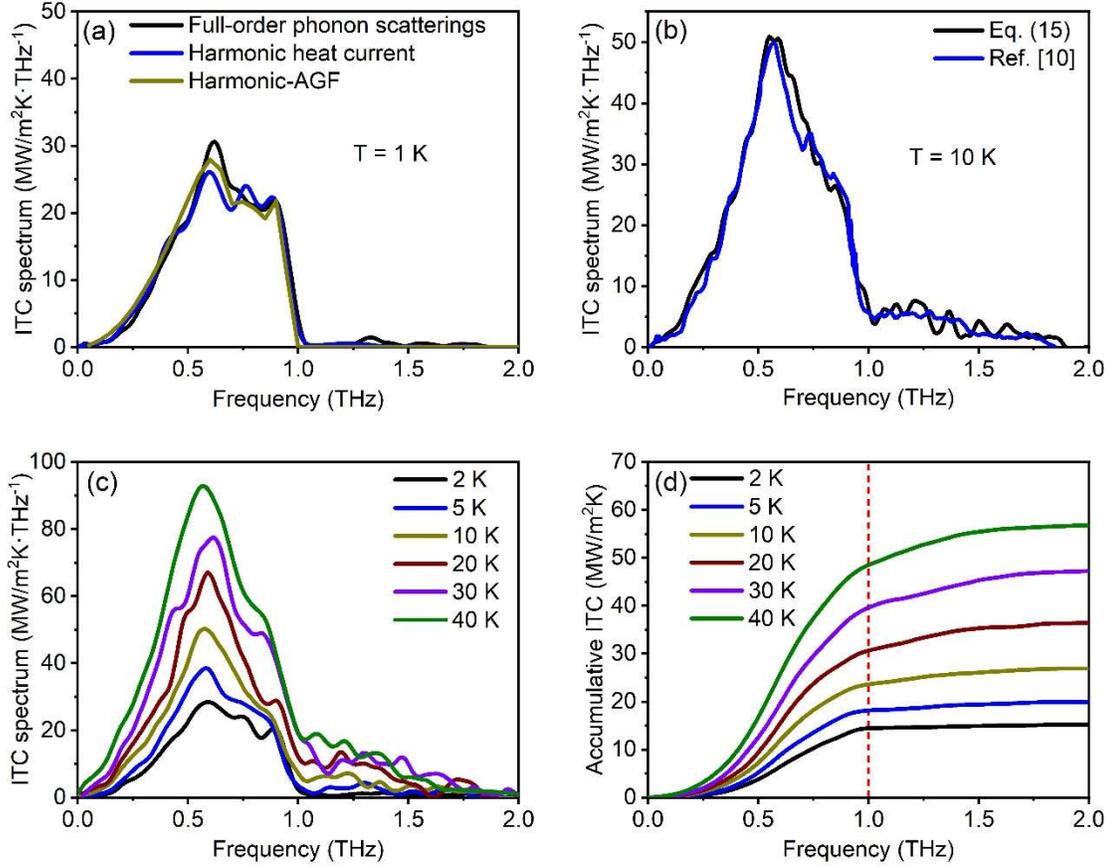

**FIG. 2** (a) The spectral thermal conductance considering the full-order phonon scatterings [Eq. (15)] at 1 K, harmonic heat current based conductance [Eq. (22)] at 1 K, and the AGF-based thermal conductance [Eq. (26)]. (b) The thermal conductance spectrum calculated via Eq. (15) at 10 K and the corresponding existing result [10]. (c) Temperature effects on the ITC spectrum at different temperatures. (d) The corresponding accumulative thermal conductance at various temperatures.

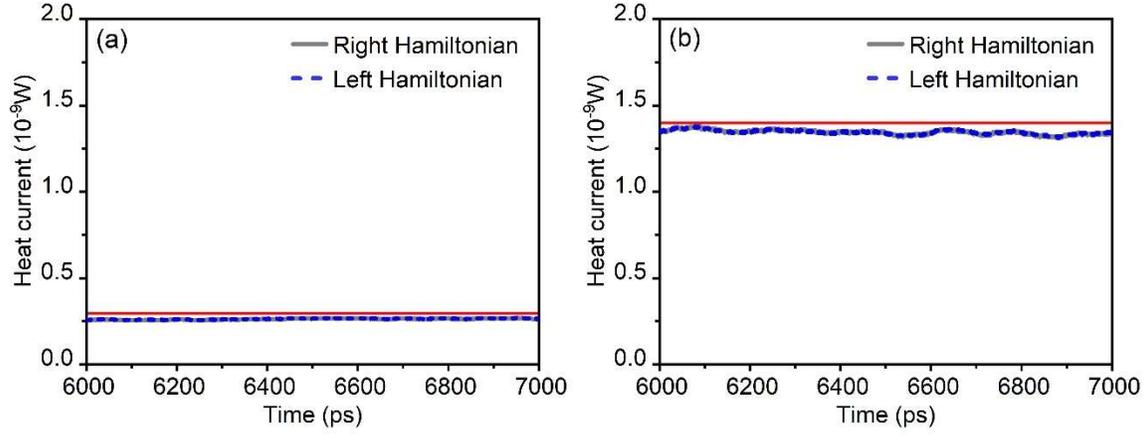

**FIG. 3** The time-dependent heat current contributed by the left and right interfacial Hamiltonians from 6 to 7 ns at two typical temperatures, i.e., (a) 2 K and (b) 10 K. $Q_{Left-H\ or\ Right-H}(t)$ is calculated using Eqs. (16) or (17). The red lines refer to half of the total heat current $Q_{NEMD}$ which is calculated via $Q_{NEMD} = \partial E_{bath} / \partial t$.

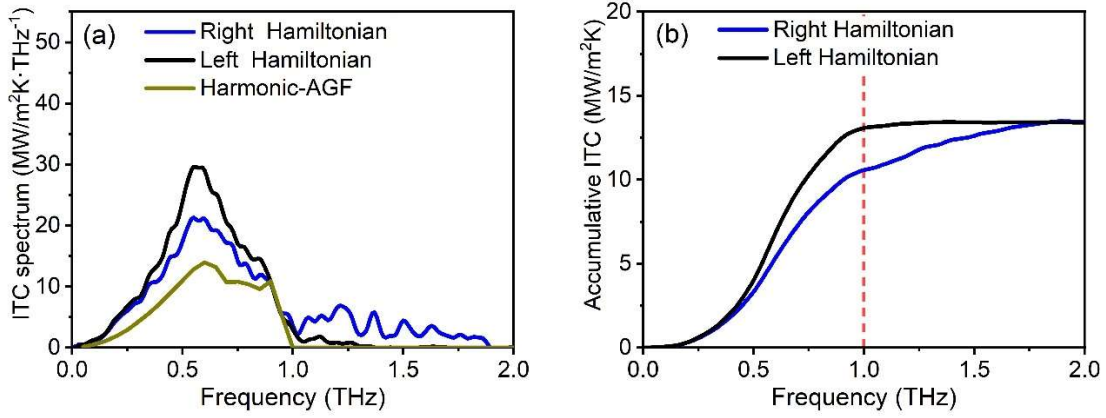

**FIG. 4** (a) The thermal conductance spectrum resulted from the left [Eq. (18)] and right [Eq. (19)] interfacial Hamiltonians at 10 K. The harmonic thermal conductance is calculated using atomistic Green's function [Eq. (26)]. (c) Accumulative thermal conductance resulted from the left and right interfacial Hamiltonians at 10 K temperature.

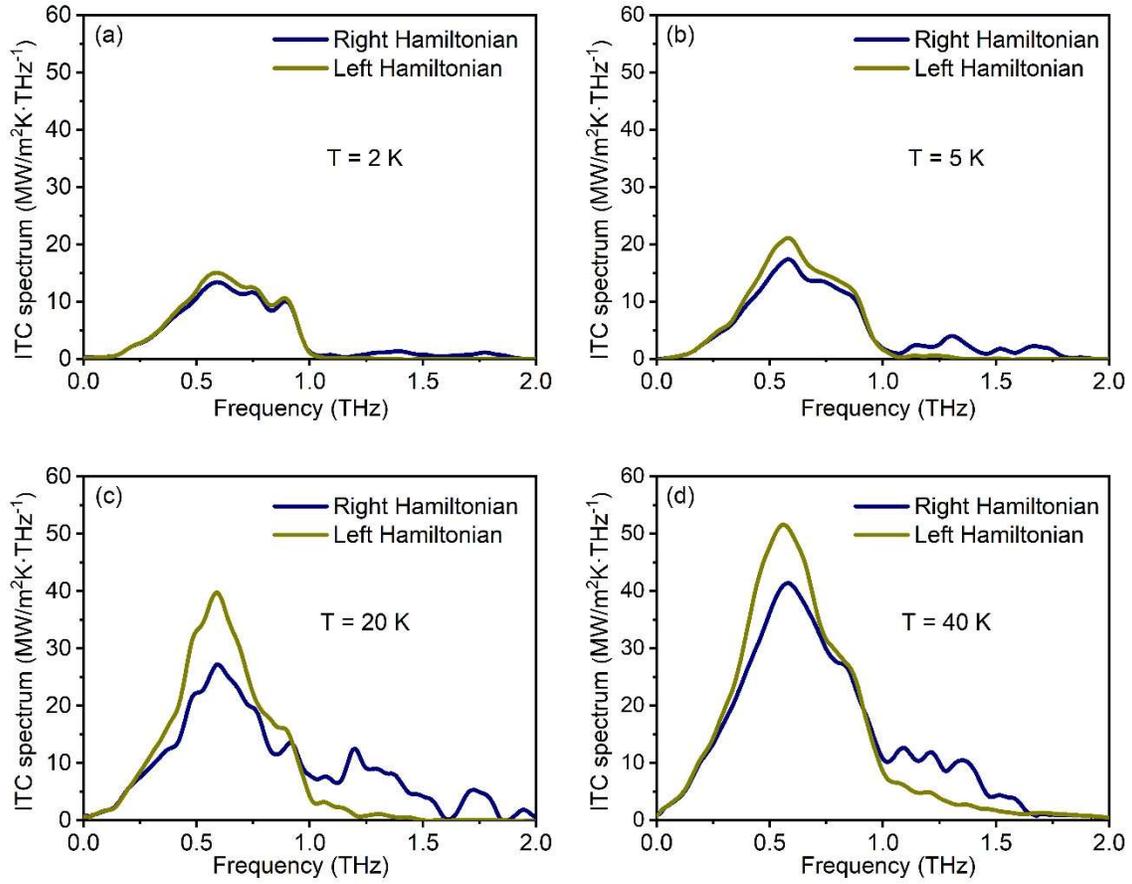

**FIG. 5** The interfacial thermal conductance spectrum resulted from the right and the left Hamiltonians at different temperatures: (a) 2 K, (b) 5 K, (c) 20 K, and (d) 40 K.

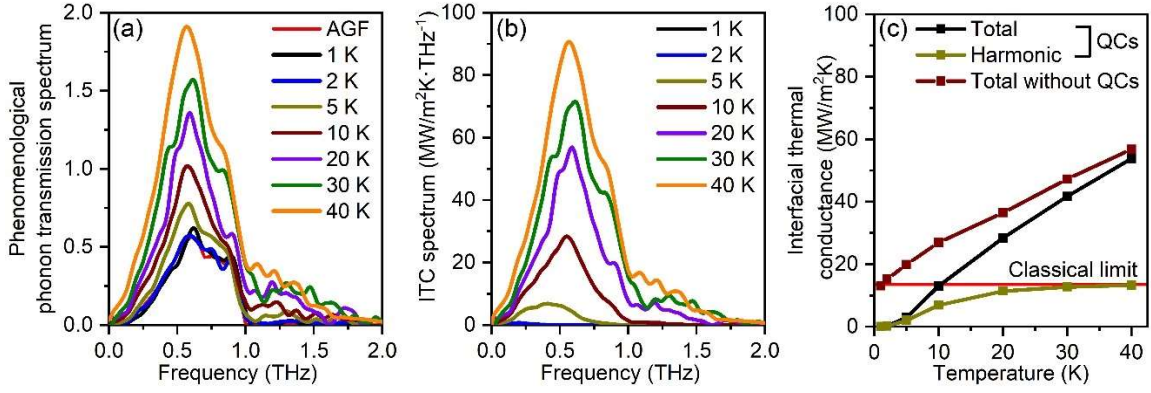

**FIG. 6** (a) The temperature-dependent spectral phenomenological phonon transmissions calculated using Eq. (27). The phonon transmission spectrum calculated using atomistic Green's functions is also plotted for comparison. (b) The corrected thermal conductance spectrum via applying the Bose-Einstein distribution in our calculations. (c) The total and harmonic thermal conductance at various temperatures with and without quantum corrections (QCs).